# Enhancing cardiovascular risk prediction through AI-enabled calcium-omics

Ammar Hoori, PhD,[1] Sadeer Al-Kindi, MD, PhD,[2,3] Tao Hu,[1] Yingnan Song,[1] Hao Wu,[1] Juhwan Lee, PhD,[1] Nour Tashtish, MD,[2] Pingfu Fu, PhD,[4] Robert Gilkeson, MD, PhD,[2,5] Sanjay Rajagopalan, MD, PhD,[2,3] David L. Wilson, PhD,[1,6,*]

[1]Department of Biomedical Engineering, Case Western Reserve University, Cleveland, OH, 44106, USA
[2]Harrington Heart and Vascular Institute, University Hospitals Cleveland Medical Center, Cleveland, OH, 44106, USA
[3]School of Medicine, Case Western Reserve University, Cleveland, OH, 44106, USA
[4]Department of Population and Quantitative Health Sciences, Case Western Reserve University, Cleveland, OH, 44106, USA
[5]Department of Radiology, University Hospitals Cleveland Medical Center, Cleveland, OH, 44106, USA
[6]Department of Radiology, Case Western Reserve University, Cleveland, OH, 44106, USA
*dlw@case.edu*

## Abstract

**Background.** Coronary artery calcium (CAC) is a powerful predictor of major adverse cardiovascular events (MACE). Traditional Agatston score simply sums the calcium, albeit in a non-linear way, leaving room for improved calcification assessments that will more fully capture the extent of disease.

**Objective.** To determine if AI methods using detailed calcification features (i.e., calcium-omics) can improve MACE prediction.

**Methods.** We investigated additional features of calcification including assessment of mass, volume, density, spatial distribution, territory, etc. We used a Cox model with elastic-net regularization on 2457 CT calcium score (CTCS) enriched for MACE events obtained from a large no-cost CLARIFY program (ClinicalTrials.gov Identifier: NCT04075162). We employed sampling techniques to enhance model training. We also investigated Cox models with selected features to identify explainable high-risk characteristics.

**Results.** Our proposed calcium-omics model with modified synthetic down sampling and up sampling gave C-index (80.5%/71.6%) and two-year AUC (82.4%/74.8%) for (80:20, training/testing), respectively (sampling was applied to the training set only). Results compared favorably to Agatston which gave C-index (71.3%/70.3%) and AUC (71.8%/68.8%), respectively. Among calcium-omics features, numbers of calcifications, LAD mass, and diffusivity (a measure of spatial distribution) were important determinants of increased risk, with dense calcification (>1000HU) associated with lower risk. The calcium-omics model reclassified 63% of MACE patients to the high risk group in a held-out test. The categorical net-reclassification index was NRI=0.153.

**Conclusions.** AI analysis of coronary calcification can lead to improved results as compared to Agatston scoring. Our findings suggest the utility of calcium-omics in improved prediction of risk.

## Introduction

There is a great need for precision risk tools to guide personalized prevention strategies for heart health. While cardiovascular risk can be estimated using many widely available cardiovascular risks scores from clinical factors, most scores suffer from poor discrimination [1]. The CT calcium score (CTCS) imaging exam can provide direct evidence of coronary atherosclerosis when calcifications are present in the coronary arteries and is acknowledged by several guidelines as a preferred risk assessment tool [2], [3]. The presence of coronary artery calcium (CAC) is by far the best predictor of future major adverse cardiovascular events (MACE) outperforming every other risk factor and composite clinical risk scoring approaches. The addition of CAC score to traditional risk factors has been shown to consistently improve discrimination and reclassification [4]. Despite their acknowledged superiority over current risk assessment approaches, current approaches for CAC-based risk prediction are overly simplistic and suffer from a number of limitations. The Agatston method simply uses a non-linear weighted sum of the areas of coronary artery calcium (CAC) with densities above 130 HU. A calcium mass score is known to be more reproducible [5]. Importantly, current CAC scoring approach ignores a plethora of other CAC features that may be pathophysiologically important, including density, distribution, geometry, and others. Some alternatives have appeared in the literature (e.g., spatial distribution, diffuse CAC, and high-density calcified plaque [6]–[8]) but never in combined fashion. Other pathophysiologic observations on calcifications have suggested a number of aspects that could be important but are currently not incorporated.

In this work, we evaluate a novel machine approach that includes features associated with calcifications (calcium-omics) and evaluate their collective ability to predict MACE in time-to-event models. Calcium-omics includes combinations of shape, mass, density, volume, number of calcifications, diffusivity, and others which together could potentially better capture a patient's risk of a MACE event.

## Methods

**Study data.** Non-contrast CTCS images were acquired from a variety of CT scanners using 120-kVp, nominally 30-mAs, with an average 0.5×0.5-mm in-plane voxel spacing and 2.5-mm slice thickness. A typical CTCS volume consists of 40 slices of 512x512 voxels, giving 10.5 million voxels per volume. We used CTCS images from 2457 patients (single CTCS volume per patient) enriched for MACE (13.8%), with characteristics in Table 1. MACE was defined as first event of myocardial infarction, stroke, coronary revascularization, or all-cause mortality. Cardiovascular outcomes were obtained from the UH CLARIFY study with a maximum of six years of follow-up (mean follow time is 1.9 years). Patient's MACE-free time is reported as the duration from the start time (time of CTCS exam) until the patient either had MACE or was censored (left the study or survived to the end of observation without MACE). This study on de-identified data was approved by the Institutional Review Board (IRB) of the University Hospitals Cleveland Medical Center.

### Image analysis and risk prediction methods

**Data preparation.** Previously, patient images were analyzed using semi-automated commercial software. The criteria for calcification detection were according to prior standards endorsed by guidelines which specified three connected voxels with HU≥130 [9]. Analysts went through each volume, slice-by-slice, and assigned each coronary calcification to a territory. For each heart, the software created a mask volume, identifying the calcifications in each territory with a different color and computed whole heart as well as territorial Agatston score. We excluded cases that had (1) poor image quality and (2) showed >10% Agatston score difference between commercial and automated in-house deep learning software. As a preprocessing step, the color-coded masks were deciphered to obtain the proper territory, creating a clean mask volume. This step required special processing to ignoring extra text labels embedded in the image. The pipeline of our proposed model is shown in Fig. 1.

**Calcium-omics feature engineering.** Using the mask volume as a guide, we created software to compute various calcium-omics features for each individual calcification, artery territory, and whole heart. For each individual calcification, we collected elemental features including mass, volume, territory, HU values, first moment, second moment, shape, distance to a subsequent lesion, distance to the top of the CT volume, artery diffusivity, among others. (Artery diffusivity was the ratio of number of calcified lesions to the Euclidian distance from first to last lesion within an artery) and represents the distribution of lesions within artery. For a territory with no calcifications or a single calcification, we set diffusivity to 0 and 1, respectively. In addition, additional statistical features such as mean, standard deviation, skewness, kurtosis, and small histogram were obtained per territory and for the entire heart. In total, we collected 80 calcium-omics features. Agatston, mass, and volume were obtained at the level of individual calcification, coronary territory, and whole heart levels. As demonstrated in Fig. 2, different features were aggerated within three levels (lesion calcification, artery, and whole heart). Details of calcium-omics features, and time-to-event modeling are described in the supplemental file.

**MACE risk prediction and performance evaluation.** We randomly divided data into training/held-out-testing subsets with 80:20 ratio for all our experiments, maintaining a similar MACE-event ratio for training and testing sets. We used the natural logarithm function to condense features with broad-range values (e.g., Agatston and mass scores). Starting with 80 calcification features, we eliminated 19 irrelevant or highly autocorrelated features by univariate Cox modeling, leaving 61 features (Table S1).

To determine high risk features and enhance explainability of results, we investigated selected univariate and multivariable Cox models. We evaluated the impact of mass scores using multivariable Cox models and investigated the impact of adding features such as the number of lesions, max HU, distance-based along territories, and CAC distribution along territories (diffusivity) to the mass score model. As a machine learning technique, Cox modeling provides interpretable results that can explain the effect of those imaging features which would be unavailable in deep learning. To enhance comparisons, all Cox models in our study were trained and tested on the same data.

We selected the most informative and non-correlated features using elastic-net as implemented in R package $glmnet$. Elastic-net was performed on the training subset using 10-fold cross-validation, with $\alpha = 0.05$, and $\lambda = 0.074$, where these parameters were determined in preliminary evaluations. Out of the 61 engineered features, elastic net selected 39 features with non-zero coefficients ($\beta$). Features include whole heart features (e.g., mass score, volume score, and number of lesions), territorial features (e.g., mass score in LAD, number of lesions in RCA, and distance from top to last lesion in LCX), and calcification features (e.g., mass histogram bin, the maximum first momentum value of individual calcification, and the third skewness value of individual calcification), as shown in table S1. These features were aggregated into a single "calcium-omics" feature by summing the products of these features by their corresponding coefficients. We used R 4.2.1 [10], the Cox model package $coxph()$, and elastic-net package $glmnet()$.

To evaluate the performance of those models, we utilized multiple time-to-event analyses. Standard metrics included C-index, AUC at fixed time points, and log-rank score. Hazard ratios with confidence intervals are presented so as to isolate the impact of a single feature. In addition, we stratified risk groups and created Kaplan-Meier (KM) plots. We also computed categorical net reclassification improvement. In some instances, we compared groups using student's t-test, with significant differences identified when p < 0.05.

## Results

Histograms of selected calcium-omics features are shown in Fig. 3. Distributions for MACE and no-MACE have considerable overlap, eliminating the possibility of creating clear-cut decision rules for MACE based on single feature, with the exception of zero total calcium score. The lack of clear discriminating thresholds suggests the need for an AI approach using multiple features at once.

We investigated multiple univariate and multivariable Cox models to understand and explain the role of particular features on MACE prediction (Table 2). Comparing Agatston (line 1) and mass score (line 2), we determined that mass score had a slightly higher C-index and AUC at two years. As the mass score is generally considered more reproducible than Agatston [5], [11], we used it in subsequent evaluations. When we examined territorial mass scores (line 3), we found improved discrimination (C-index and AUC) compared to a whole heart mass score particularly for the LAD, which has the lowest p-value and the highest HR, indicating that an equivalent mass in the LAD was more predictive of MACE than that in another territory. For a given mass score, increasing the number of lesions had a significant effect (p<0.004) (line 4). For every unit increase of (ln(1+NumLesions)), the risk of having MACE increased by 1.48-fold. Hence, compared to one lesion, the risk was increased by 227% for 40 lesions, a number sometimes observed. Adding a logical feature to indicate two or more territories with calcification did not improve mass score model (p=0.12) (line 5 versus line 2). In contrast, for a given mass score, HU >1000 was protective (line 6). The distance from the "top" to "bottom" calcification per territory (line 7) improved performance with regard to log-rank score, C-index, and AUC as compared to other models (lines 1-6), even though no single territory produced a significant effect (p<0.05). The number of lesions per total distance in each territory (diffusivity as described in Methods) performed better than lesion distance (line 8 versus line 7). Regarding their HRs, diffusivity in territories ranked as LM > RCA > LCX > LAD. A calcium-omics model with 39 features after elastic net regularization (line 9) was highly predictive of MACE with (HR = 3.62, p<0.0001). When compared with other features (lines 1-8), the calcium-omics model had the best performance metrics in multiple categories. With the use of sampling to improve the event rate and elastic-net determination of 59 features, our calcium-omics model with sampling (line 10) yielded even better performance. Compared to the conventional standard (Agatston score, line 1) on held-out test data, this model improved C-index from 70.3% to 71.6% and the year-2 AUC from 68.8% to 74.8%. As these metrics are notoriously difficult to improve, we deem this increase substantive.

In Fig. 4, we show year-2 ROCs for the calcium-omics model with and without sampling and compare them to the conventional Agatston score. Without sampling, calcium-omics gave (training/testing) AUCs of (74.7% / 71.4%), while the Agatston model gave (71.8% / 68.8%), respectively. Utilizing modified-SMOTE sampling, the calcium-omics model gave AUCs of (82.4% / 74.8%), while sampling did not affect the Agatston score model. Similarly, at year-3, calcium-omics with sampling gave the best results. However, at year-3, there were fewer cases due to censoring and events, giving more uncertain results.

The Agatston model has a number of limitations. In contrast, the calcium-omics model is more discriminating due to its capacity to accommodate a broader range of calcium-omics features. While whole Agatston score had a non-linear relationship with MACE events in the log hazard ratio regression curve (Fig 5), the calcium-omics model had a more linear curve. These curves were plotted using the Cox model of penalized spline of a feature and calculated the log of hazard ratio of each patient to show the distribution along the regression curve. The calcium-omics model showed a wide range of risk levels for cases with similar Agatston scores in an interactive 2D surface regression plot (right plot in Fig. 5) implying good distinguishable values for cases having similar Agatston score. The contours display areas that result in equivalent levels of severity. Interestingly, the plot shows the capability of calcium-omics to cover a wide variation of values for narrow Agatston score values.

Consistent with the linear relationship between the calcium-omics model and MACE shown in Fig 5, calcium-omics stratified risk groups better than did Agatston (Fig. 6). For the Agatston score (left), patients were stratified into the five risk groups recognized by the Lipid Association with Agatston score ranges (0, 1-99, 100-299, 300-999, and 1000+) [12], [13]. For calcium-omics, we set thresholds for the aggerated calcium-omics feature that gave the same proportions of patients in the risk groups as in the Agatston score plot. The calcium-omics model separated risk groups much better than does the Agatston model. Focusing on year two for Agatston, the three middle-risk curves are not very informative giving nearly the same MACE-free proportions. For calcium-omics at this time, there is informative, good separation. These results are consistent with the left and middle log hazard ratio regression curves in Fig. 5.

Calcium-omics improved net reclassification in the held-out test set. Figure 7 shows Kaplan-Meier curves for the 20% held-out-test subset. As this smaller data set is insufficient to support 5-group stratification, we use a single cut-off of Agatston=100, a value often considered in the literature. As before, the threshold for creating the calcium-omics curves gave the same starting patient proportions as for Agatston. Again, there is better separation provided by the calcium-omics model than the Agatston model. Considering results at year four, the calcium-omics model was able to reclassify 63% of patients with MACE and 35% of patients without MACE who were considered in the wrong risk group (within study period) for Agatston score model. The overall categorical net reclassification improvement was, $NRI = 0.154\ [95\%\ CI\ 0.006 - 0.302; p = 0.042]$, indicating improvement in the proposed model.

Figure 8 highlights the limitation of the whole-heart Agatston score in two patients who have approximately equal Agatston scores (~204), but one has diffuse disease with 11 lesions in three territories, and the other has only two lesions in one territory. Whole heart Agatston would have predicted the same risk, but our calcium-omics approach predicts that at three years, the patient with only two lesions (right) will have a MACE-free survival probability 2.3 times better than the other patient with diffuse disease (left). The high-risk patient had a MACE later in the study period.

## Discussion

In this paper we provide an initial evaluation of an integrated radiomic approach that incorporates 80 different features spanning multiple elemental features of shape, texture, distribution and statistical parameters to predict MACE and compared this with the traditional Agatston score. The use of calcium-omics was far more discriminatory than Agatston, improving the AUC from 68.8 to 74.8 (p=0.07). The relationship between an aggregated calcium-omics score and MACE was nearly linear with a graded effect, when compared to Agatston which displayed a non-linear relationship particularly at high levels of calcium score (Fig. 5). Importantly, calcium-omics resulted in a graded dose effects on MACE, as opposed to considerable overlap in risk across risk stratification quartiles (Fig. 6). While our findings are intuitive in the sense that incorporation of multiple features may be expected to enhance risk prediction, this is not always true given that many features may be correlated.

The success of calcium-omics relative to Agatston lies in its ability to better characterize coronary artery disease as compared to the Agatston score, which is simply a summation of calcium in the coronaries, albeit in a non-linear way. Calcium-omics captures characteristics from individual calcifications, including mass, volume, HU values, numbers of calcifications, numbers of territories, and spatial distribution. Univariate and multivariable Cox models in Table 2 offer explanation as to why calcium-omics does better, per the following observations. 1) Summing over the entire heart, mass score slightly improves prediction compared with Agatston, potentially due to its improved reproducibility [5], [11]. 2) When we simply add the dense calcification (>1000 HU) feature to mass, there is improvement as compared to mass or Agatston alone. Highly calcified "older" lesions are likely more stable explaining this finding [8]. 3) Adding mass scores from individual arteries improves performance. This suggests that having disease present in more than one artery is a risk factor. This is directly shown as a logical (≥2) in line 5 giving HR=1.45 after accounting for total mass score. 4) Adding the number of calcifications to the whole heart mass score greatly improved risk prediction, again suggesting that the spread of disease is a risk factor. 5) Our diffusivity metric is a risk factor indicating that the spread of disease along arteries is a risk factor. Taking all features together in calcium-omics simply provides more information about the disease, enabling improved overall risk prediction.

There are important contributions of our work. Calcium-omics outperforms the current state of the art – whole heart Agatston score. This development could contribute to more precise personalized therapies for cardiovascular patients. We purposefully used a MACE-enriched cohort in this, our first study on AI analysis of calcium-omics features. This smaller cohort allowed us to carefully vet all data to ensure data quality. Overall performance might be different when larger numbers of cases with a low event rate are used in future studies. We found that modified-SMOTE up sampling and down sampling reduces the problem with low event rate data. This is the first time that such strategies have been used with CT calcium score data. Often overlooked, down sampling and up sampling has been previously described for coronary heart disease cohort [14].

Our study undoubtedly has limitations. Importantly, we had a limited observation period (average 1.9 years within the 6-years study), which limits event rates. The data used in this study were from sites across the University Hospitals Health System, which is restricted to northeast Ohio. Other locales might have somewhat different results. Additionally, data used in this study were obtained using various scanners with similar acquisition parameters. We did not perform analyses to identify model performance by scanner type.

In conclusion, we have obtained promising results using an AI analysis on detailed calcification features. Clearly, there will be advantage as compared to the standard whole-heart Agatston score. It is hoped that results will carry over to larger, confirming studies.

**Declaration.** Human subject research has been done under an IRB of Case Western Reserve University and University Hospitals Cleveland Medical Center. Images were acquired at University Hospitals Cleveland Medical Center and shared under a data use agreement. This research was supported by National Heart, Lung, and Blood Institute through grants R01HL167199, R01HL165218, R01 HL143484, R44HL156811, and a pilot study from award 1P50MD017351-01. The content of this report is solely the responsibility of the authors and does not necessarily represent the official views of the National Institutes of Health. The grants were obtained via collaboration between Case Western Reserve University and University Hospitals Cleveland Medical Center. This work made use of the High-Performance Computing Resource in the Core Facility for Advanced Research Computing at Case Western Reserve University.

# Figure and tables titles and legends

| Characteristic | Full Cohort | No-MACE | MACE |
|---|---|---|---|
| Patients | 2457 | 2118 (86.2%) | 339 (13.8%) |
| Zero Ag Score | 957 (38.9%) | 942 (44.5%) | 15 (4.4%) |
| Female | 1185 (48.2%) | 1044 (49.3%) | 141 (41.6%) |
| Age* | 60.7±9.6 (19, 90) | 60±9.5 (19,90) | 65.6±8.5 (41,87) |
| Time within study* | 699 (3, 2192) days  1.9 (0, 6) years | 726 (31, 2192)  2 (0, 6) | 530.5 (3,2170)  1.5 (0, 5.9) |
| Agatston Score* | 220± 364.4(0, 1992) | 190.9±337.2 (0, 1992) | 402.2±462.6 (0, 1969) |
| Mass Score* | 33.9±56.4 (0, 342) | 29.4±51.7 (0, 327.8) | 62.4±73.6 (0, 342) |
| Volume Score* | 186.3±299.2 (0, 1819.7) | 161.3±275.7 (0, 1819.7) | 342±382.1 (0, 1776.8) |
| Num Lesions* | 5±6.9 (0,65) | 4.5±6.4 (0, 46) | 9±8.1 (0, 65) |
| * Numbers reported as mean ± standard deviation (min, max) values. | | | |

**Table 1**. Characteristics of our randomly chosen cohort of 2457 enriched with regards to MACE events. Characteristics are given for the full cohort and the MACE and no-MACE groups. The cohort has great variability along clinical features (female and gender) high percent of zero Agatston score (38.9%). The image-driven score features such as Agatston, mass, and volume score, in addition to the total number of lesions, show good statistical distributions. MACE vs. no-MACE sub-cohorts shows great stratification along all features.

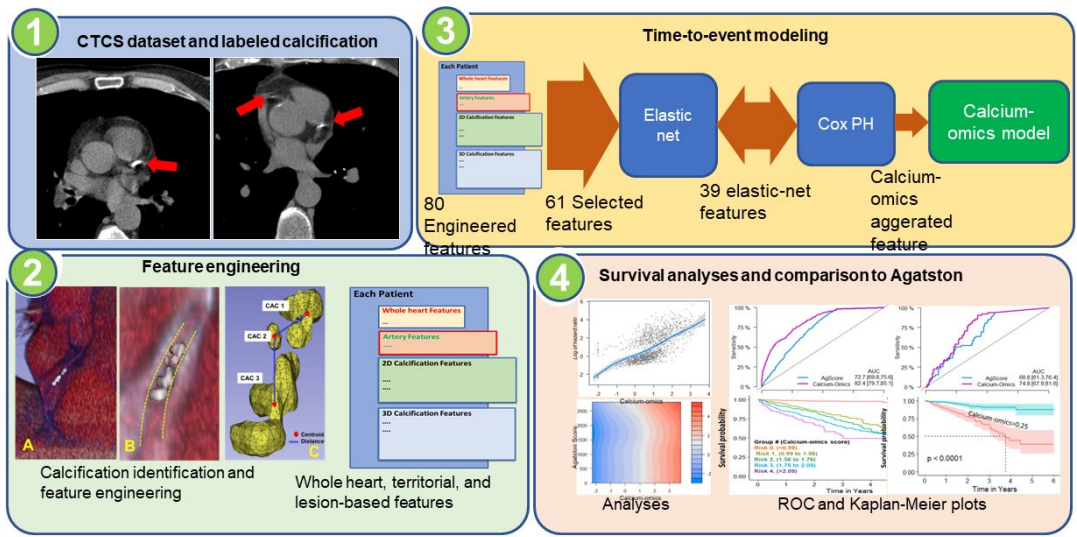

**Figure 1.** MACE prediction using calcium-omics features pipeline. In (1), CAC lesions in CTCS images were analyzed and labeled using semi-automated commercial software. In (2), calcium-omics features were engineered and categorized based on whole heart, territorial, and lesion features. In (3) MACE risk prediction model was designed using elastic-net and Cox model. In (4), results and statistical analyses were performed to assess the importance of our novel calcium-omics model compared to variety of univariate and multivariable Cox models.

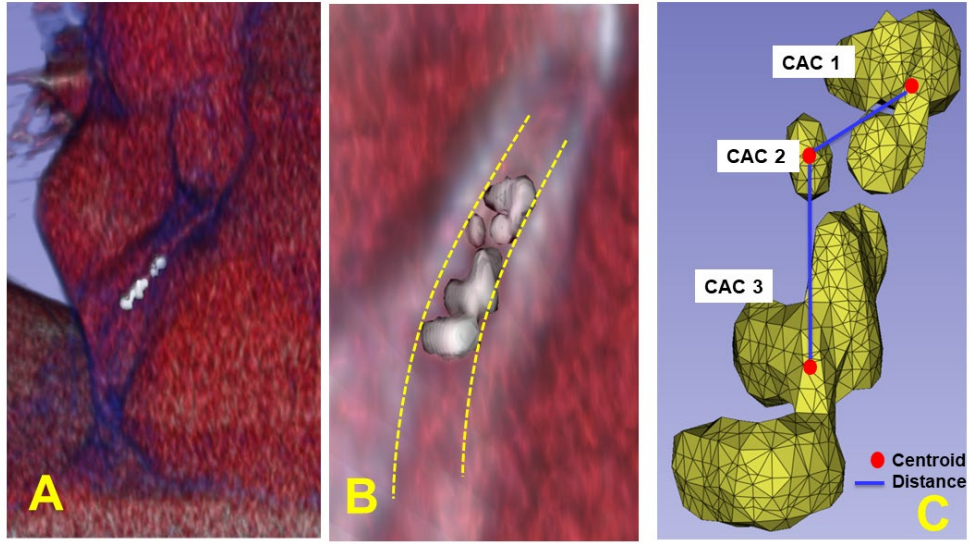

**Figure 2.** Individual calcifications and engineered calcium-omics features. Three consecutive calcifications in the LCX artery territory are shown in (A) and magnified in (B), where dashed lines annotate the vessel wall. Calcification masks are rendered in (C). Some features are aggregated along each artery, such as Agatston, mass, and volume scores, which give this LCX artery 84.4, 13.3, and 73.2, respectively. Calcification centroids are used to calculate the Euclidean distances between calcifications. The sub-voxel centroid (x, y, z) locations are used to calculate the calcified arterial distance to sum DistFirst2LastLesionPerArtery from a centroid to a centroid in consecutive sequential order. An example of a new feature, is "DistFirst2LastLesionPerArtery_LCX" which represents the total Euclidean distance along lesions within LCX.

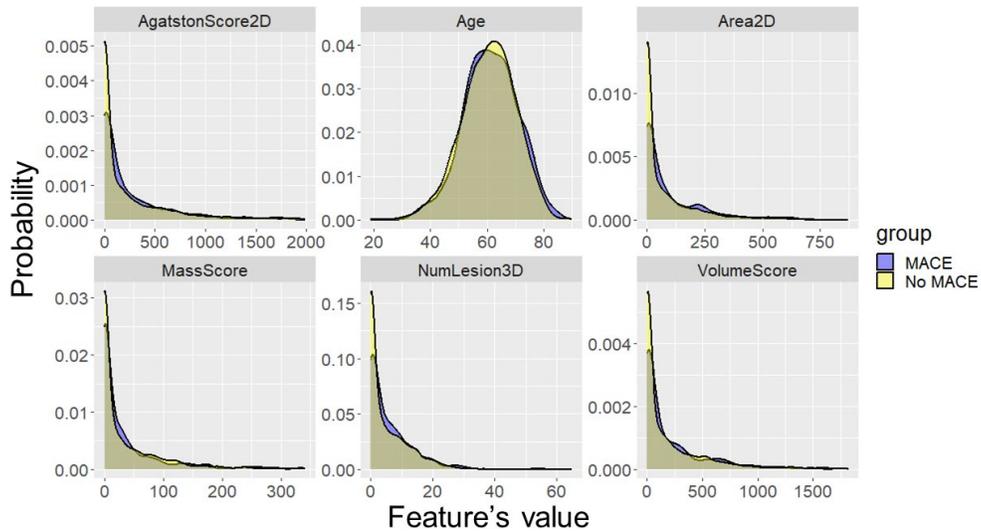

**Figure 3.** Normalized histograms of feature values for the MACE and no-MACE groups. For the 80 features we analyzed, we found that no single feature, including Agatston and mass scores, gave strong visual evidence of differences between groups. However, because of the large number of samples, t-tests gave $p<0.0001$, allowing us to reject the null hypothesis of no difference in the means. Of course, these histograms do not consider censoring as done with time-to-event modeling. The x-axis represents the values of each feature, while the y-axis is the probability of histogram bins as obtained by normalizing the histogram.

| Cox PH Model | Cox feature(s) | HR [±95% CI; p-value] | Log-rank score | C-index (train) % | C-index (test) % | AUC (train) % | AUC (test) % |
|---|---|---|---|---|---|---|---|
| 1. Agatston score | Ln(1+Agatston Score) | 1.39 [1.31,1.47; <0.0001]*** | 157.4 | 71.3 | 70.3 | 71.8 | 68.8 |
| 2. Mass score | Ln(1+Mass Score) | 1.48 [1.38, 1.59; <0.0001]*** | 144.8 | 71.1 | 70.6 | 71.6 | 68.9 |
| 3. Arterial mass scores | Ln(1+LM Mass)<br>Ln(1+LAD Mass)<br>Ln(1+LCX Mass)<br>Ln(1+RCA Mass) | 1.13 [1.0,1.27; 0.034]*<br>1.27 [1.15,1.4; <0.0001]***<br>1.12 [1.0,1.25; 0.039]*<br>1.1 [1.0,1.2; 0.062] | 138.6 | 71.2 | **72.1** | 71.8 | 69.3 |
| 4. Number of lesions | Ln(1+Mass Score)<br>Ln(1+NumLesions) | 1.2 [1.03,1.41; 0.021]*<br>1.48 [1.34,1.92; 0.004]** | 151.9 | 71.6 | 71.0 | 71.9 | 70.4 |
| 5. Number of calcified arteries | Ln(1+Mass Score)<br>Is_ClacifedArteries>=2 | 1.39 [1.23,1.55; <0.0001]***<br>1.45 [0.91,2.31; 0.120] | 145.2 | 71.3 | 70.8 | 71.6 | 69.7 |
| 6. HUmax >=1000 | Ln(1+Mass Score)<br>is_HUmaxAbove1000 | 1.53 [1.42,1.67; <0.0001]***<br>0.71 [0.51,0.99; 0.042]* | 147.4 | 71.2 | 70.4 | 72.1 | 69.0 |
| 7. Lesions' distance (Top to last lesion) | Ln(1+Mass Score)<br>LM distance<br>LAD distance<br>LCX distance<br>RCA distance | 1.38 [1.23, 1.54;<0.0001]***<br>1.001 [0.996,1.007; 0.617]<br>1.000 [0.997,1.003; 0.998]<br>1.003 [0.999,1.006; 0.102]<br>1.001 [0.999,1.003; 0.396] | 156.5 | 71.7 | 71.3 | 72.4 | 69.7 |
| 8. Territorial diffusivity | Ln(1+Mass Score)<br>LM diffusivity<br>LAD diffusivity<br>LCX diffusivity<br>RCA diffusivity | 1.43 [1.32, 1.56; <0.0001]***<br>2.89 [1.05, 7.98; 0.041]*<br>1.2 [0.26, 5.63; 0.815]<br>1.52 [0.52, 4.43; 0.449]<br>2.2 [0.4, 10.3; 0.310] | 154.2 | 71.8 | 71.6 | 72.5 | 70.0 |
| 9. Calcium-omics | Calcium-omics (39 features) | 3.62 [2.92,4.48; <0.0001]*** | **185.9** | **74.0** | 70.8 | **74.6** | **71.4** |
| **10. Calcium-omics (with sampling)** | Calcium-omics (59 features) | 2.81 [2.6,3.03; <0.0001]*** | **808** | **80.5** | 71.6 | **82.4** | **74.8** |

Notes: (1) P-values are star-coded based on the significance levels as follows: (<0.0005 as ***, <0.005 as **, <0.05 as *).
(2) Red is the highest value, while black bolded is the second highest value.

**Table 2.** Comparison of calcification risk models. To explain the role of particular features, especially high-risk features, we investigated multiple univariate and multivariable Cox models. Rows are models with different features or feature subsets. Columns are self-explanatory. We include results on both training and held-out testing data. The p-values are used to reject the null hypothesis that HR=1 in the Cox model. See text for a detailed analysis of results.

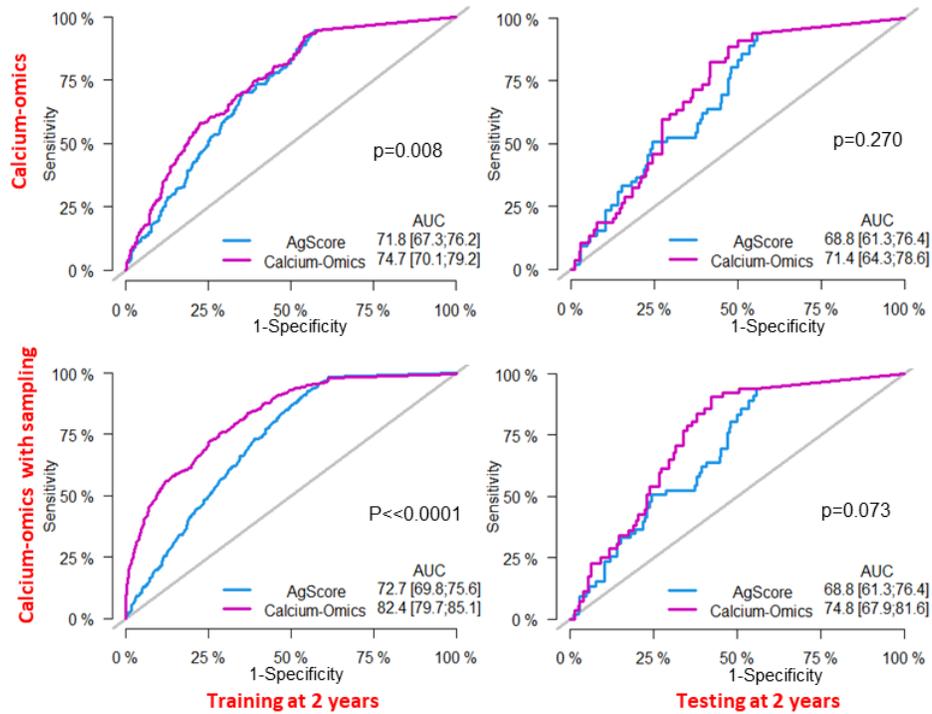

**Figure 4.** Performance of calcium-omics risk prediction with and without sampling (bottom and top, respectively). Along each row, calcium-omics ROCs are shown for training at 2 years, and testing at 2 years, respectively. Agatston score results are shown for comparison. Across the board, calcium-omics was superior to Agatston. Calcium-omics performance was improved with sampling and yielded a significant difference to Agatston (p<<0.0001) as compared to no sampling calcium-omics to Agatston (p=0.008). For sampling, we used a modified-SMOTE (see text) with down sampling and up sampling on training data only. The held-out test set was not subjected to any data sampling strategy. The p-values correspond to the Wald test for the AUC significance of a model compared to the rival model.

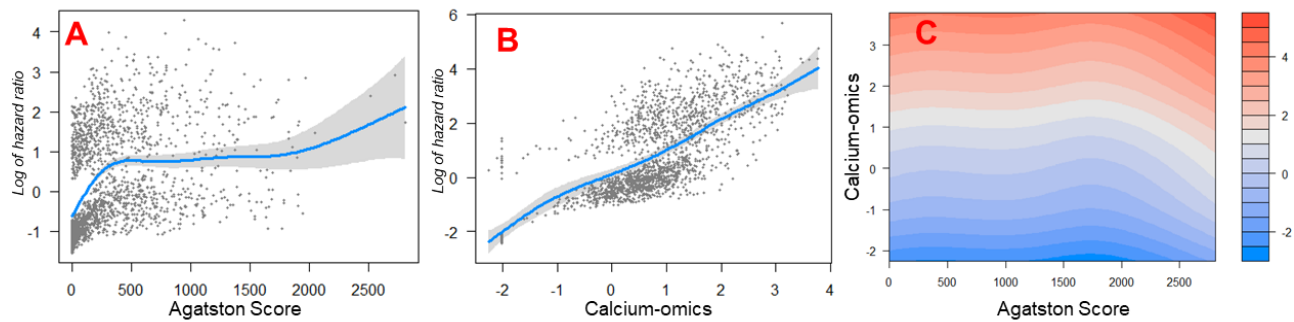

**Figure 5.** Log hazard ratio, i.e., ln[h(t)/h$_0$(t)], regression plots for Cox models as a function of Agatston (A) and calcium-omics scores (B). Visualizations are available using the *visreg( )*, and *visreg2D( )* functions in *visreg* R Library. Briefly, we used penalized spline (*pspline*) function to create the blue log hazard ratio curves. Each data point represents a patient's deviance residual, and the shaded-gray areas show the 95% CI. As compared to the Agatston model, the calcium-omics model shows a desirable, linear distribution along the data points. In the case of the Agatston model (A), a wide range of Agatston score (300-2000) gives very similar results. Similar observations are shown in (C), where the log hazard ratio is displayed in gradient-colored contours, from low (blue) to severe (red). Calcium-omics is platted as a function of Agatston. At a given value of Agatston, there

is considerable variation of the log hazard calculated from calcium-omics. For example, with an Agatston score of 500, several levels of severity are covered by the calcium-omics model. This suggests added value.

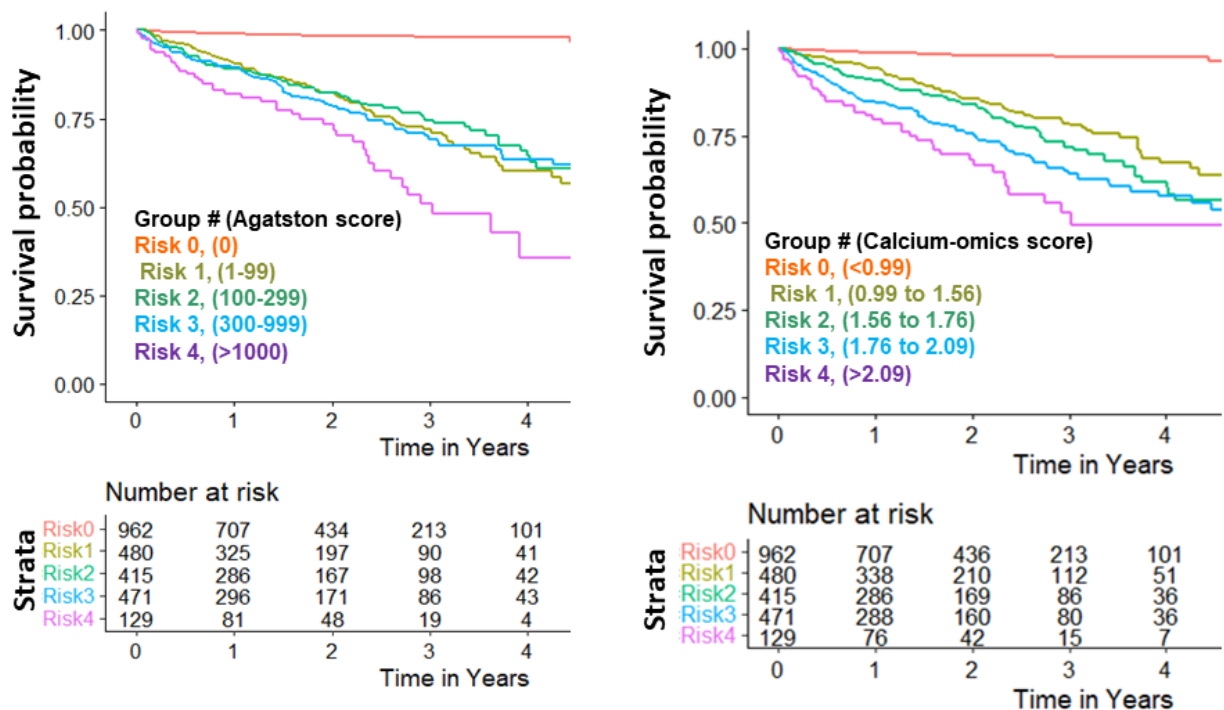

**Figure 6.** Kaplan Meier survival (MACE-free) curves with stratification provided by Cox modeling for all data. Plots represent the full MACE-enriched cohort using the standard Agatston score model (left) vs. calcium-omics model (right). The x-axis represents survival time, while the y-axis represents the survival probability of patients within a risk group. Agatston score was stratified into 5 groups according to the Lipid Association recommendation with Agatston score ranges (0, 1-99, 100-299, 300-999, and >1000). Groups for the calcium-omics model were created with scores (<0.99, 0.99-1.56, 1.56-1.76, 1.76-2.09, >2.09) to give equivalent numbers of patients as for Agatston. The five risk severity groups are ordered (0-4), where 0 is the lowest-risk. Visually, the calcium-omics model much better stratified the five groups as compared to Agatston. In particular, risk groups 1, 2, and 3 are much more clearly separated for calcium-omics than Agatston. Due to the low number of held-out test samples, these plots are done with all data.

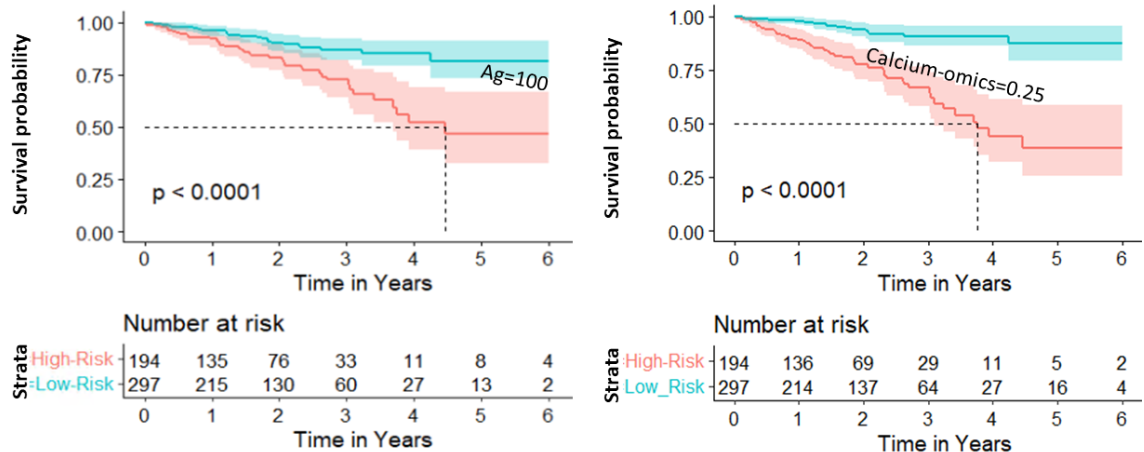

**Figure 7.** Kaplan Meier survival (MACE-free) curves with stratification provided by Cox modeling for held-out test set. Agatston score model stratified into low-risk (cyan) and high-risk (pink) risk groups based on below or above Agatston score of 100 (Left). In a similar ratio to the left model, the calcium-omics model stratified patients into low and high-risk (right) with a calcium-omics feature value =0.25. The calcium-omics model showed better visual separable stratification by reclassifying some patients to fit into high or low-risk groups. Survival probability of 50% was reached at year 4.5 with Agatston model, while reached in 3.8 years in calcium-omics, showing advantageous to the latter model. At year four, we investigated the calcium-omics model categorical reclassification performance compared to Agatston score model. For the patients with MACE, the calcium-omics model showed a categorical net reclassification improvement of $NRI_{MACE} = 0.132$. Calcium-omics was able to correctly reclassify 63% and miss reclassify 20%. While with No-MACE patients, the new model showed $NRI_{NoMACE} = 0.022$, and was able to correctly reclassify 35% and miss reclassify 17%. The total NRI showed advantage to the new model with $NRI_{Total} = 0.154\ [95\%\ CI\ 0.006,\ 0.302;\ p = 0.042]$.

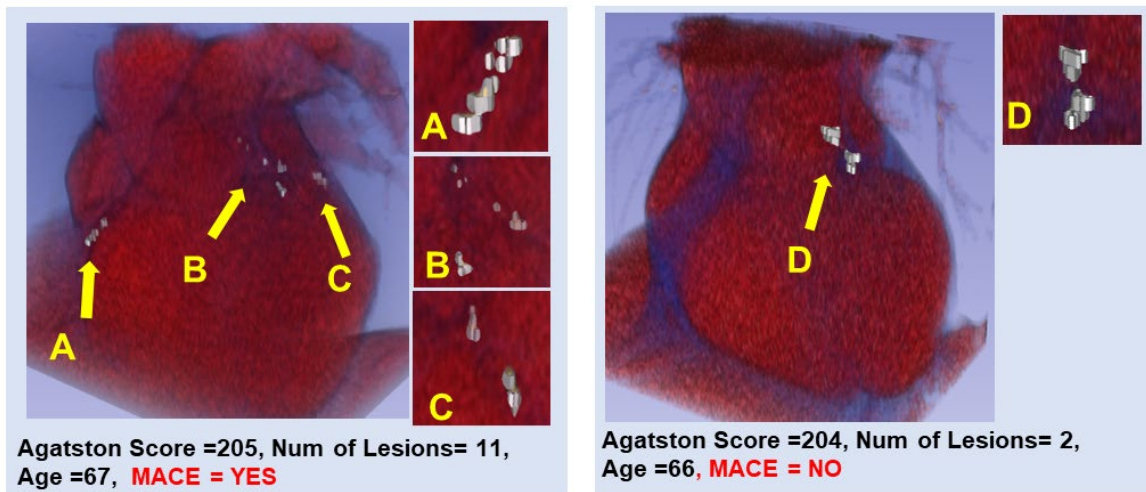

**Figure 8**. Whole-heart Agatston does not reflect the spread of disease and risk for these two patients, both with a whole-heart Agatston score of ~204. The left heart has 11 calcifications spread throughout the heart (i.e., LAD:6, LCX:3, and RCA:2), with Agatston scores of (29.6, 84.4, and 90.8), respectively. The right heart has two "nearby" large calcifications (LM:1, LAD:1) with Agatston scores of (108.2 and 95.6), respectively. Both patients are from the held-out test set, with the same Age (~67). Despite an equal whole-heart Agatston score, the calcium-omics model described later predicted a 3-year risk for the left heart 2.3 times that of the right heart. The patient on the left patient had a MACE event, while the right did not have MACE.

## Supplemental Materials:
## S.1 Detailed feature engineering

The three main traditional whole-heart scores (Agatston, mass, and volume) are given below.

***1. Agatston score.*** Agatston score uses a weighting factor depending upon the maximum HU value for each lesion in a 2D CT image. The total Agatston score was calculated by summing the product of the density weighting factor (DWF) and the 2D area of each calcified lesion. (Some larger 3D lesions will have multiple 2D entries.) The whole heart Agatston score is obtained as given below.

$$Agatston\ score = \frac{3}{ST} \sum_{i=1}^{N} DWF(MaxHU_i) \times Area_i \qquad (1)$$

Here, $N$ is the number of 2D lesions, $MaxHU_i$ is the maximum HU value within the $i$-th lesion. $Area_i$ is the $i$-th lesion 2D area in $mm^2$. As the Agatston score was originally calculated with a $3\ mm$ slice thickness, we adjust values with the ratio, $3/ST$, where ST is the new slice thickness. Values for $DWF()$ are given below [19].

$$DWF(x) = \begin{cases} 1, & 130 \leq x < 200 \\ 2, & 200 \leq x < 300 \\ 3, & 300 \leq x < 400 \\ 4, & 400 \leq x \end{cases} \qquad (2)$$

***2. Mass score.*** The absolute mass of coronary calcium score is aggregated per lesion with the aid of phantom calibration and evaluated as follows:

$$Mass\ score = \sum_{i=1}^{N} \sum_{v=1}^{M} k \times HU_v \qquad (3)$$

where $k$ is a calibration factor converting HU to mg. $N$ is the number of lesions. $M$ is the number of voxels within lesion $i$, and $HU_v$ represents the HU value of the selected voxel $v$. For a Philips scanner at 120 kVp, $k$ = 0.71 $(mgHA/cm^3)/HU$.

***3. Volume score:*** Volumes were obtained by simply summing labeled voxels $v = [1,..,M]$, and multiplying the volume per voxel, $V$.

$$Volume\ score = VM, \qquad (4)$$

In addition to whole heart aggregated features (such as Agatston score, Volume score, and mass score), we collected lesion, lesion-to-lesion, and arterial-wise features. We calculated per artery score features, including Agatston score, mass score, and volume score. We engineered more calcium-driven features such as lesion aggregated areas, HU statistical features (min, max, average, median, and standard deviation), distance from the first slice to last calcification, and distance from first to last lesion along descending arterial lesions. We also collected lesion-based statistical histogram bins of the first moment, second moment, mean moment, skewness moment, kurtosis moment, and average HU. Some of these features are briefly explained as follows, where 2D features are slice-based, and 3D features are volume-based:

1-***Numerical*** features include:
- **Area2D** (total heart summation of lesions' areas across all slices)
- **NumLesion3D** (total heart number of 3D lesions)
- **numLesionPerArtery3D_<<*name*>>1** (num of 3D lesion in specified artery)
- **AgatstonScore2D** (heart total Agatston score calculated in slice-based lesions, original Dr. Agatston approach)
- **AgatstonScore3D** (heart total Agatston score calculated in 3D volume-based lesions)
- **AgatstonScorePerArtery2D_<<*name*>>1** (<<*name*>> artery Agatston score calculated in slice-based lesions)
- **MassScorePerArtery_<<*name*>>1** (<<*name*>> artery mass score)
- **VolumeScorePerArtery_<<*name*>>1** (<<*name*>> artery volume score)
- **massHist<<number>>** (histogram bin <<number>> out of 5 bins of lesions-based mass score)
- **avrHist<<number>>** (histogram bin <<number>> out of 5 bins of mean HU values)

- **DistTop2LastLesionPerArtery_<<name>>1** (Euclidean distance summation in mm, starting from center of top CT slice along centroid of each consecutive lesion till last lesion within <<name>> artery)
- **DistFirst2LastLesionPerArtery_<<name>>1** (Euclidean distance summation in mm, starting from centroid of first lesion, along centroid of each consecutive lesion till last lesion within <<name>> artery)
- **ICfirstMomentH<<number>>** (max values of first momentum among individual calcifications, order <<number>>) (<<number>> up to 3 values)
- **ICsecondMomentH<<number>>** (max values of second momentum among individual calcifications, order <<number>>) (<<number>> up to 3 values)
- **ICmeanMomentH<<number>>** (max values of mean momentum among individual calcifications, order <<number>>) (<<number>> up to 3 values)
- **ICskewnessMomentH<<number>>** (max values of skewness momentum among individual calcifications, order <<number>>) (<<number>> up to 3 values)
- **ICkurtosisMomentH<<number>>** (max values of kurtosis momentum among individual calcifications, order <<number>>) (<<number>> up to 3 values)
- **HUperArtery2D_stat<<name>><<number>>** (<<number>> [1-4] represents [min, max, mean, std] statistical values of Hounsfield Units of each calcified voxel within artery<<name>>)
- **<<name>>_diffus** (factor indicates diffusivity of lesions within <<name>> artery, calculated as the ratio of number of lesions to Euclidean distance along lesions with artery from first to last lesion. We considered the non-calcified artery to have zero diffusivity while the single lesion artery to have diffusivity=one)

2-*Categorical and conditional* (Boolean) features include:
- **isAgZero** (is Agatston score equal zero?)
- **isLesion3DBelow5** (is number of lesions less than 5?)
- **AgGroupX1-X3**, Agatston score groups of (0,1-99, 100-399, 400+) represented in three (X1, X2, X3) Boolean digits to be used in Cox.
- **isArt2plus** (are there two or more calcified arteries?)
- **isArt3plus** (are there three or four calcified arteries?)
- **numArtCalc** (number of calcified arteries 0-4)
- **HU1000** (Does the patient have any calcified lesions with HU value above 1000?)

These image-based engineered features are listed in Table S1. We exclude features that are clinical or highly correlated. Among the remaining 61 features, an elastic net with 10-fold cross-validation selected 40 features, as indicated, with their corresponding Cox model coefficient values. The elastic-net Cox proportional hazard model was deemed the Calcium-omics model.

*Selected 61 engineered features (elastic-net selected only 39, shown with coefficients)*

| Feature | coef | Feature | coef | Feature | coef | Feature | coef |
|---|---|---|---|---|---|---|---|
| MassScore* | -0.6812 | VolumeScore* | -0.0774 | Area2D* | 1.21 | NumLesion3D* | -0.9914 |
| isAgZero | - | isLesion3DBelow5 | - | AgGroupX1 | - | AgGroupX2 | - |
| AgGroupX3 | - | numLesionPerArtery3D_LM1* | 0.07381 | numLesionPerArtery3D_LAD1* | - | numLesionPerArtery3D_LCX1* | - |
| numLesionPerArtery3D_RCA1* | 0.1398 | isArt2plus | - | isArt3plus | - | numArtCalc | -0.1718 |
| AgastonScorePerArtery2D_LM1* | 0.1083 | AgastonScorePerArtery2D_LAD1* | -0.2756 | AgastonScorePerArtery2D_LCX1* | -0.1445 | AgastonScorePerArtery2D_RCA1* | -0.07467 |
| MassScorePerArtery_LM1* | - | MassScorePerArtery_LAD1* | 0.4396 | MassScorePerArtery_LCX1* | 0.3636 | MassScorePerArtery_RCA1* | 0.1652 |
| VolumeScorePerArtery_LM1 | - | VolumeScorePerArtery_LAD1 | 0.00048 | VolumeScorePerArtery_LCX1 | -0.00101 | VolumeScorePerArtery_RCA1 | - |
| massHist1 | 2.332 | massHist2 | 2.152 | massHist3 | 0.7786 | massHist4 | 2.521 |
| massHist5 | 1.366 | avrHist1 | 0.7336 | avrHist2 | 0.1187 | avrHist3 | -0.4072 |
| avrHist4 | - | avrHist5 | 0.8539 | DistTop2LastLesionPerArtery_LM1 | - | DistTop2LastLesionPerArtery_LAD1 | - |
| DistTop2LastLesionPerArtery_LCX1 | 0.00679 | DistTop2LastLesionPerArtery_RCA1 | 0.00358 | DistFirst2LastLesionPerArtery_LM1 | - | DistFirst2LastLesionPerArtery_LAD1 | - |
| DistFirst2LastLesionPerArtery_LCX1 | 0.00062 | DistFirst2LastLesionPerArtery_RCA1 | - | ICfirstMomentH1 | -0.01407 | ICfirstMomentH2 | -0.00435 |
| ICfirstMomentH3 | -0.0073 | ICsecondMomentH1 | 4.9E-05 | ICsecondMomentH2 | - | ICsecondMomentH3 | - |
| ICmeanMomentH1 | 0.04454 | ICmeanMomentH2 | - | ICmeanMomentH3 | - | ICskewnesstMomentH1 | -0.01654 |
| ICskewnesstMomentH2 | 0.00288 | ICskewnesstMomentH3 | -0.0002 | ICkurtosisMomentH1 | 0.3444 | ICkurtosisMomentH2 | 0.02595 |
| ICkurtosisMomentH3 | -0.3646 | | | | | | |

*Excluded features before using elastic net*

| Feature | Feature | Feature | Feature |
|---|---|---|---|
| AgatstonScore2D* | AgatstonScore3D* | HUperArtery2D_stat_LM2 | HUperArtery2D_stat_LM3 |
| HUperArtery2D_stat_LM4 | HUperArtery2D_stat_LAD2 | HUperArtery2D_stat_LAD3 | HUperArtery2D_stat_LAD4 |
| HUperArtery2D_stat_LCX2 | HUperArtery2D_stat_LCX3 | HUperArtery2D_stat_LCX4 | HUperArtery2D_stat_RCA2 |
| HUperArtery2D_stat_RCA3 | HUperArtery2D_stat_RCA4 | LM_diffus | LAD_diffus |
| LCX_diffus | RCA_diffus | HU1000 | |

* This feature was represented in logarithmic function as log(x+1)

**Table S1.** List of all selected and excluded (before using elastic-net) image-based calcification-driven features. 19 features were excluded prior to the proposed model design due to their high correlation with other features. We used them in designing univariate and multivariable Cox models to investigate and compare with other models (such as Agatston score, HU1000, and LAD_diffus). Among the 61 listed features, an elastic net with 10-fold cross-validation selected 39 features, as indicated, with their corresponding Cox model coefficient values. The elastic net Cox proportional hazard model was deemed the calcium-omics model. These features were used to design the calcium-omics Cox model using training data without sampling.

## S. 2 Time-to-event modeling with Cox proportional hazard model and elastic-net regularization

For a clinical study at a fixed time with persons entering at various times, censoring of the observation time is an issue requiring time-to-event modeling rather than binary classification. A time-to-event model estimates the probability that the event (MACE in our study) may have occurred during a follow-up period. Whether the patient had an event or being censored, data can be modeled by a distribution function [20] of observed time $T$, at a patient survival time $t$, called the cumulative incidence function:

$$F(t) = P(T < t) = \int_t^\infty f(u)du, \qquad (5)$$

where $f(t)$ is the probability density function. $P(T < t)$ is the probability function that survival time is less than $t$. The survival function $S(t)$, is $1 - F(t)$, which is the probability that the time $T$ is greater or equal to $t$.

$$S(t) = P(T \geq t) = 1 - F(t). \qquad (6)$$

The hazard function is represented as the risk of hazard of an event occurring at time $t$ and is defined as:

$$h(t) = \frac{f(t)}{S(t)}. \qquad (7)$$

The Cox proportional-hazard regression model [21] is widely used in survival modeling. The Cox model provides a semi-parametric hazard rate of each covariate in the model, as follows:

$$h(t|A) = h_0(t)\exp(\beta_1 a_1 + \beta_2 a_2 + \cdots + \beta_n a_n), \qquad (8)$$

where $h_0(t)$ is the baseline hazard, $A = [a_1, a_2, \ldots a_n]$ is the covariate feature vector of $n$ features, $\beta_i$ is the $i_{th}$ covariate coefficient. The Cox model is optimized using maximum-likelihood. We used Cox regression for univariate and small multivariable models to study feature effects and identify high risk features. There are practical considerations. So as not to over-emphasize large covariate values, we compress dynamic range by taking a logarithm of some covariates (e.g., log (Agatston Score)). As Cox modeling is sensitive to correlated features, results may not reflect the actual effect of one feature over another. Too many features can result in over-fitting. We used elastic-net regularization with cross-validation [22] to select the best features.

To overcome the effect of low event rates [18], we applied down sampling followed by up sampling techniques on the majority and minority class, respectively. We used a modified Synthetic Minority Over-sampling Technique (modified-SMOTE) approach. For major class down sampling, we used few continuous features (e.g., Agatston score, mass score, and volume score) to determine eligibly samples to be removed using k-nearest neighbors (KNN) (k=5) in feature space. For up-sampling, we created synthetic instances "nearby" actual samples in "covariate space." Briefly, we used similar features (as in down sampling), synthetic instance was inserted within KNN (k=5). For the new sample, continuous feature value was calculated as the median of the corresponding k-neighbors feature value, while non-continuous (logical and categorical) feature values were copied from the nearest single neighbor. The new instance time-to-event was randomly set. Down sampling was done until MACE events increased from 13.8% in the original data to 16.4% by removing 20% of the No-MACE cases. Followed by up sampling, new cases were inserted until the MACE events increased to 30%. We never applied up or down-sampling on held-out test data.